\documentclass[prd,showpacs,amsmath,amssymb,superscriptaddress,nofootinbib,showkeys]{revtex4}
\usepackage{amssymb}
\usepackage{amsmath}
\usepackage{graphicx}

\normalbaselines
\newcommand{\Fst}{{\mathop {\rule{0pt}{0pt}{F}}\limits^{\;*}}\rule{0pt}{0pt}}

\begin{document}

\title{Nonminimal Einstein-Maxwell-Vlasov-axion model}
\author{Alexander B. Balakin}
\email{Alexander.Balakin@kpfu.ru} \affiliation{Department of
General Relativity and Gravitation, Institute of Physics, Kazan
Federal University, Kremlevskaya str. 18, Kazan 420008, Russia}

\author{Ruslan K. Muharlyamov}
\email{Ruslan.Muharlyamov@kpfu.ru} \affiliation{Department of
General Relativity and Gravitation, Institute of Physics, Kazan
Federal University, Kremlevskaya str. 18, Kazan 420008, Russia}

\author{Alexei E. Zayats}
\email{Alexei.Zayats@kpfu.ru} \affiliation{Department of General
Relativity and Gravitation, Institute of Physics, Kazan Federal
University, Kremlevskaya str. 18, Kazan 420008, Russia}

\begin{abstract}
We establish a new self-consistent system of equations accounting
for a nonminimal coupling of the cooperative gravitational,
electromagnetic and pseudoscalar (axion) fields in a
multi-component relativistic plasma. The axionic extension of the
nonminimal Einstein-Maxwell-Vlasov theory  is based on two
consistent procedures. First, we use the Lagrange formalism to
obtain nonminimal equations for the gravitational, electromagnetic
and pseudoscalar fields with the additional sources generated in
plasma. Second, we use the Vlasov version of the relativistic
kinetic theory of the plasma, guided by the cooperative
macroscopic electromagnetic, gravitational and axionic fields, to
describe adequately the response of the plasma on the variations
of these fields. In order to show the self-consistency of this
approach we check directly the compatibility conditions for the
master equations for the cooperative fields. Using these
compatibility conditions we reconstruct the ponderomotive force,
which acts on the plasma particles, and discuss the necessary
conditions for existence of the distribution function of the
equilibrium type.
\end{abstract}

\pacs{04.40.-b, 52.25.Dg, 14.80.Va}

\keywords{Axion field, relativistic plasma, Vlasov model,
nonminimal coupling}

\maketitle

\section{Introduction}

\subsection{What does the Einstein-Maxwell-Vlasov-axion model describe?}

The Einstein-Maxwell-Vlasov-axion model deals with the
self-consistent theory of interaction between gravitational,
electromagnetic, pseudoscalar (axion) fields and a relativistic
multi-component plasma. The specific concept introduced by Vlasov
into the plasma theory, namely, the concept of {\it cooperative}
macroscopic field \cite{Vlasov1,Vlasov2} (or, equivalently,
self-consistent, collective, average, mean field) had became the
highly sought one in modern physics. Let us remind, that the
well-known Maxwell-Vlasov model (the basic element of modern
plasma theory \cite{MV1,MV2}) deals with the cooperative
electromagnetic field, which is created by the electrically
charged particles in plasma and which regulates the plasma state
variations. From the mathematical point of view, the Lorentz
force, which enters the kinetic equation, contains the tensor of
electromagnetic field (the Maxwell tensor, $F_{ik}$), which is the
solution of the Maxwell equations with the electric current
four-vector, averaged over the ensemble of these charged
particles. Thus, according to the Vlasov concept, the ensemble of
plasma particles controls itself by means of cooperative
electromagnetic field. In this context, the Einstein-Vlasov model
(see, e.g., \cite{EV1,EV2}) deals with the cooperative
gravitational field, which is created by some statistical system
(e.g., gas or plasma) and guides this system in a self-consistent
manner. In the Einstein-Vlasov models with scalar field (see,
e.g., \cite{EVS1}) new (scalar) sources contribute to the
creation of gravitational field, in which the kinetic system
evolves. The axionic extension of the Vlasov idea is based on the
introduction of cooperative pseudoscalar (axion) field; this model
is not yet elaborated, here we start to discuss the Vlasov-type
models of the axionically active plasma. When we combine the
Einstein-Vlasov, Maxwell-Vlasov and axion-Vlasov models, we deal
with the cooperative gravitational-electromagnetic-pseudoscalar
field, which regulates the behavior of relativistic plasma and
displays specific cross-interactions. To illustrate the appearance
of cross-interactions let us remind, for instance, the details of
the theory of relativistic gravitationally coupled plasma systems,
studied in various contexts, e.g., in
\cite{EMV1,EMV4,EMV5}. In fact, we deal there with the
application of the Einstein-Maxwell-Vlasov model, and the coupling
of cooperative gravitational and electromagnetic fields of the
plasma is the example of such cross-interaction. Clearly, the
presented here Einstein-Maxwell-Vlasov-axion model includes the
description of the gravitational-electromagnetic,
gravitational-axion, axion-photon cross-couplings, and all these
interactions can be unified on the base of the Vlasov concept of
the cooperative field of the plasma system.

\subsection{What does the nonminimal coupling scheme add to the model?}

The description of the nonminimal coupling of the gravitational
field with scalar, pseudoscalar (axion), electromagnetic, massive
vector and gauge fields is based on the introduction of specific
cross-terms into the Lagrangian, which contain the Riemann tensor
$R_{ikmn}$, the Ricci tensor, $R_{kn}$, and Ricci scalar, $R$, on
the one hand, and the corresponding fields and their derivatives,
on the other hand. The theory of nonminimal coupling is elaborated
in detail for the scalar field (real and complex fields $\Phi$, as
well as Higgs multiplets $\Phi_{({\rm a})}$) (see, e.g.,
\cite{FaraR,HO,OdinPhRep} for review and references). Special
attention in these investigations is focused on two models, the
first of them has the $\xi R \Phi^2$ coupling, and the second has
the so-called nonminimal derivative coupling
\cite{derc1,derc2}. The study of a nonminimal coupling of
gravity with electromagnetic field started in \cite{Prasanna1} and
this theory has been developed by many authors (see, e.g.,
\cite{Drum,Go,Kost1,Hehl2,BL05,H3}).
Exact solutions in the framework of nonminimal Einstein-Yang-Mills
and Einstein-Yang-Mills-Higgs theories are discussed in
\cite{BZ1,BSZ,BDZ1}. Nonminimal models of the
axion-photon coupling were considered in \cite{BaNi}.

When we deal with the nonminimal Einstein-Maxwell-Vlasov-axion
model, we should, first, recover all the cross-terms appeared in
the nonminimal Einstein-Maxwell-axion field theory \cite{BaNi}.
The second step in this way is connected with the description of
the nonminimal coupling of plasma to gravity. In
\cite{Rmatter1,Rmatter2} one can find the models of the nonminimal
coupling of gravity to matter. We follow here the Vlasov concept
and use another way: we introduce into the kinetic equation the
tidal (curvature induced) force linear in the Riemann tensor
related to the cooperative gravitational field in analogy with the
Stokes, Langevin, antifriction, etc. forces considered in
\cite{tide2,tide5,tide10}.
Thus, in our model we describe the nonminimal coupling of gravity
with fields (electromagnetic and pseudoscalar) and with matter
(plasma particles).

\subsection{Why this model could be interesting for applications to cosmology and astrophysics?}

Our Universe expands with acceleration. This phenomenon,
discovered in the observations of the Supernovae Ia
\cite{A1,A2,A3}, can be explained in two different ways. The first
explanation is based on the hypothesis that there exists a dark
energy \cite{DE1,DE2,DE3}, a cosmic substrate with negative
pressure. Together with the dark matter \cite{DM1,DM2,DM3}, the
existence of which is usually associated with the observations of
the flat velocity curves of the spiral galaxies rotation
\cite{Sp1,Sp2}, the dark energy form the so-called cosmic dark
fluid. The total contribution of the dark energy and dark matter
into the energy balance of the Universe is estimated to be about
$95 \%$, thus the dark fluid guides the Universe late-time
evolution and predetermines its fate. The second way of the
explanation of the Universe accelerated expansion is connected
with the generalization of the Einstein gravitational theory (see
\cite{OdinPhRep} for review and references), the nonminimal field
theory being one of the versions of such generalization. Axions,
hypothetical light massive pseudo-bosons, which appeared in the
cosmological lexicon in the context of the strong CP-violation
problem and spontaneous breaking of symmetry in the early Universe
(see, e.g., \cite{Peccei,8Ni,Weinberg,Wilczek0,Peccei2,Battesti}),
are considered as one of the dark matter candidates (see, e.g.,
\cite{Raffelt,Turner,Khl} and references therein). Keeping in mind
that the contribution of the dark matter into the total Universe
energy-density is about 23\%, we should consider in detail the
role of the pseudoscalar (axion) field in the Universe evolution,
when we use the nonminimal field theory approach to its
description.

Another motif to consider the Vlasov version of the
Einstein-Maxwell-axion-plasma model is connected with the idea of
self-regulation (self-guiding) of the cosmic dark fluid based on
the Vlasov concept of a cooperative field. Such a self-regulation
might, for instance, avoid the Big Rip scenaria \cite{BR1} of the
late-time Universe evolution  in analogy with the dynamic
singularity avoidance in the $F(R)$, $f(T)$-gravity models and
their modifications, in the models with Chaplygin gas, in the
model for the dark energy with various effective time-dependent
equations of state (see, e.g., \cite{avo1,LR1}).

In addition, we keep in mind that plasma is an important
constituent of many objects and media in our Universe, and that
the photons emitted, scattered or deflected by the plasma
particles, propagate in the environment of axionic dark matter.
Thus, the minimal and nonminimal models of the axion-photon
coupling in a relativistic plasma can clarify some properties of
the electromagnetic waves emitted by astrophysical sources and
detected by astronomers. In particular, one can expect detection
of fingerprints of the axionic dark matter in the spectra of
plasma frequencies and in the polarization rotation effects
\cite{19}.

\subsection{Our goal and structure of this paper}

We are interested in establishing and detailed analysis of a new
self-consistent Einstein-Maxwell-Vlasov-axion model, which
describes evolutionary processes in a relativistic multi-component
plasma guided by cooperative
gravitational-electromagnetic-pseudoscalar (axion) fields. In this
work we formulate both minimal and nonminimal
Einstein-Maxwell-Vlasov-axion-plasma models using the combined
Lagrange formalism and the kinetic approach. In order to show
self-consistency of this combined approach we check directly the
compatibility conditions for the coupled master equations
describing the cooperative gravitational, electromagnetic and
pseudoscalar (axion) fields in plasma.

This work is organized as follows. In Section~\ref{MinEMVa} we
formulate the Vlasov approach to the description of the minimal
Einstein-Maxwell-axion-plasma model. In Subsection~\ref{MinEMVa1}
we consider the Lagrangian composed of classical terms for the
gravitational, electromagnetic, pseudoscalar fields, as well as
the term describing the minimal axion-photon coupling, and a
general additive term for matter. In Subsection~\ref{MinEMVa2} we
obtain minimal master equations for the gravitational,
electromagnetic, pseudoscalar (axion) fields with sources induced
by plasma (in general form), and the corresponding compatibility
conditions. In Subsection~\ref{MinEMVa3} we formulate the kinetic
equation of the Vlasov type (\ref{MinEMVa31}), consider the
moments of the distribution function (\ref{MinEMVa32}),
reconstruct the effective force using the compatibility conditions
(\ref{MinEMVa33}), discuss the problem of entropy balance
(\ref{MinEMVa34}), and discuss the structure of the distribution
function of the equilibrium type as a solution to the kinetic
equation with cooperative gravitational, electromagnetic and axion
fields (\ref{MinEMVa35}). Section~\ref{NMEMVa} contains the
nonminimal generalizations of the master equations for the
gravitational, electromagnetic and axion fields, as well as the
analysis of the generalized compatibility conditions. In
Section~\ref{Concl} we discuss the obtained results.
Appendices~\ref{AppA} and \ref{AppB} contain auxiliary and
preparatory formulas, which are necessary for the self-consistent
formulation of the model.

\section{Minimal Einstein-Maxwell- Vlasov-axion
model}\label{MinEMVa}

\subsection{Action functional}\label{MinEMVa1}

We start with the action functional
\begin{gather}
S_{({\rm M})} {=} \int d^4 x \sqrt{{-}g} \left\{
\frac{R{+}2\Lambda}{2\kappa}
 {+} \frac{1}{4} F^{mn} \left(F_{mn}{+}\phi \Fst_{mn}\right){+}
 L_{({\rm m})} {-}
\frac{1}{2}\Psi^2_0 \left[g^{mn} \nabla_m \phi \nabla_n \phi {-}
m^2_{({\rm A})} \phi^2 {-} V({\phi}^2) \right] \right\},
\label{R1}
\end{gather}
where $g_{ik}$ is the spacetime metric, $g$ is its determinant,
$\nabla_k$ denotes the covariant derivative, $R$ is the Ricci
scalar, $\Lambda$ is the cosmological constant, $\kappa$ is the
Einstein constant. The Maxwell tensor $F_{mn}$ is expressed in
terms of four-vector potential of the electromagnetic field $A_m$:
\begin{equation}
F_{mn} \equiv \nabla_m A_{n} - \nabla_n A_{m} \,.
\label{1maxtensor}
\end{equation}
The term $\Fst^{mn} \equiv \frac{1}{2} \epsilon^{mnpq}F_{pq}$
describes the tensor dual to the Maxwell tensor; $\epsilon^{mnpq}
\equiv \frac{1}{\sqrt{-g}} E^{mnpq}$ is the Levi-Civita tensor,
$E^{mnpq}$ is the absolutely antisymmetric Levi-Civita symbol with
$E^{0123}=1$. The dual Maxwell tensor satisfies the condition
\begin{equation}
\nabla_{k} \Fst^{ik} =0 \,. \label{Emaxstar}
\end{equation}
The symbol $\phi$ is used for the dimensionless pseudoscalar field; the axion field itself, $\Phi$, is considered to
be proportional to this quantity $\Phi {=} \Psi_0 \phi$ with a coupling
constant $\Psi_0$. The term $m_{({\rm A})}$
is proportional to a (hypothetical) mass of an axion, $m_{({\rm
A})} =  \frac{m_{({\rm axion})}}{\hbar}$; $\hbar$ is the Planck
constant; $V({\phi}^2)$ is a potential of the pseudoscalar field. We use the signature $+---$, and the units with $c{=}1$.
The term $L_{({\rm m})}$ stands for the Lagrangian of a matter interacting in general case with electromagnetic, pseudoscalar and gravitational fields; this term is considered to be
the function of $A_i$, $\phi$ and $g_{ik}$.

\subsection{Master equations for the cooperative fields}\label{MinEMVa2}

\subsubsection{Electrodynamic equations}

The variation of the action functional (\ref{R1}) with respect
to the four-vector potential $A_i$ gives the electrodynamic equations
\begin{equation}
\nabla_k \left( F^{ik} + \phi \Fst^{ik}\right)= - \frac{\delta
L_{({\rm m})}}{\delta A_i} \,, \label{R4}
\end{equation}
which can be transformed into
\begin{equation}
\nabla_k F^{ik} = - \Fst^{ik} \nabla_k \phi - I^i \,,
\label{E2177}
\end{equation}
using the equations (\ref{Emaxstar}) and the standard definition of the electric current four-vector $I^i$, induced in plasma:
\begin{equation}
\frac{\delta L_{({\rm m})}}{\delta A_i} \equiv  I^i \,.
\label{R2}
\end{equation}
The term $\Fst^{ik} \nabla_k \phi$ in (\ref{E2177}) can be indicated as axionically induced effective current.

\subsubsection{Master equation for the pseudoscalar (axion) field}

The variation procedure with respect to the pseudoscalar field $\phi$  yields
\begin{equation}
\left[\nabla^k  \nabla_k {+} m^2_{({\rm A})} {+} V^{\prime}(\phi^2)
\right] \phi = {-} \frac{1}{\Psi^2_0} \left[\frac{1}{4}
\Fst^{mn}F_{mn} {+} {\cal J} \right]  , \label{induc3}
\end{equation}
where the following definition is used for the scalar ${\cal J}$:
\begin{equation}
{\cal J} \equiv \frac{\delta L_{({\rm m})}}{\delta \phi}   \,. \label{induc03}
\end{equation}
Two source-like terms in the right-hand side of this equation
appear: the first relates to the standard electromagnetic source
for the axion field, the second one can be interpreted as a scalar
current provided by the interaction of plasma particles with the axion
field.

\subsubsection{Master equations for the gravitational field}

The variation of the action functional (\ref{R1}) with respect to the metric $g^{ik}$ gives the gravity field equations
\begin{equation}
R_{ik} - \frac{1}{2}Rg_{ik} = \Lambda g_{ik} + \kappa \left[ T^{({\rm EM})}_{ik} + T^{({\rm A})}_{ik} +
T^{({\rm m})}_{ik} \right] \,. \label{011}
\end{equation}
Here the stress-energy tensor of the electromagnetic field is of the standard form
\begin{equation}
T^{({\rm EM})}_{ik} \equiv  \frac{1}{4}g_{ik} F_{mn}F^{mn} - F_{im} F_{k}^{\ m} \,, \label{1TEM}
\end{equation}
the stress-energy tensor of the pseudoscalar field is
\begin{gather}
T^{({\rm A})}_{ik} \equiv \Psi^2_0 \left\{\nabla_i \phi \nabla_k
\phi -  \frac{1}{2} g_{ik} \left[ \nabla^m \phi \nabla_m \phi -
m^2_{({\rm A})} \phi^2 - V(\phi^2)\right] \right\} \,, \label{TAX}
\end{gather}
and the stress-energy tensor of the plasma particles is given by the following expression:
\begin{equation}
T^{({\rm m})}_{ik} = - \frac{2}{\sqrt{-g}} \frac{\delta \left[\sqrt{-g} L_{({\rm m})}\right]}{\delta g^{ik}} \,. \label{1EineqMIN}
\end{equation}
Let us stress that the variation of the term
\begin{equation}
\frac{1}{4} \sqrt{{-}g} \ \phi F^{mn} \Fst_{mn} = \frac{1}{8} \phi
E^{ikmn}F_{ik}F_{mn} \label{1axi}
\end{equation}
vanishes, since this term does not depend on metric.

\subsubsection{Compatibility conditions}

Two compatibility conditions are well-known; the first one
\begin{equation}
\nabla_k I^k = 0 \,,
\label{2axi}
\end{equation}
is the differential consequence of the equation (\ref{E2177});
the second condition follows from the Bianchi identities and yields
\begin{equation}
\nabla^k \left(R_{ik} - \frac{1}{2}Rg_{ik}\right) \equiv 0 =
\kappa \nabla^k \left[ T^{({\rm EM})}_{ik} + T^{({\rm A})}_{ik} +
T^{({\rm m})}_{ik} \right] \,. \label{3axi}
\end{equation}
Using the equations (\ref{E2177}) and (\ref{induc3}) one can rewrite the last condition in the form
\begin{equation}
\nabla^k T^{({\rm m})}_{ik} = F_{ik} I^k + {\cal J} \nabla_i \phi \,. \label{4axi}
\end{equation}
Here we used the formula
\begin{equation}
F^i_{\ m}\Fst^{km} = \frac{1}{4}g^{ik} \Fst_{mn}
F^{mn}\,,\label{5axi}
\end{equation}
which allows us to reduce the pseudotensor $F^i_{\ m}\Fst^{km}$
(see Appendix A for details). Now we have to divide the
stress-energy tensor of the matter $T^{({\rm m})}_{ik}$ into two
parts
\begin{equation}
T^{({\rm m})}_{ik} = {\cal T}^{({\rm m})}_{ik} + T^{({\rm D})}_{ik} \,.
\label{totalT1}
\end{equation}
The first part, ${\cal T}^{({\rm m})}_{ik}$, relates to the
(massive) plasma particles, and this stress-energy tensor will be
represented below in terms of kinetic theory as a second-order
moment of the distribution function. The term $T^{({\rm D})}_{ik}$
corresponds to nonkinetic substrates, for instance, in the
cosmological context it describes a dark fluid composed of a dark
energy and dark matter. We suppose that $T^{({\rm D})}_{ik}$ can
be written in the standard form
\begin{equation}
T^{({\rm D})}_{ik} = \rho^{({\rm D})}U_i U_k +I^{({\rm D})}_i U_k + I^{({\rm D})}_k U_i + {\cal P}^{({\rm D})}_{ik} \,,
\label{totalT2}
\end{equation}
where $\rho^{({\rm D})}$ is the energy-density (e.g., of the dark
fluid), $I^{({\rm D})}_i$ is the heat-flux four-vector, $U^k$ is
the velocity four-vector of the substrate, and ${\cal P}^{({\rm
D})}_{ik}$ is the pressure tensor.

\subsection{Kinetic representation of the material sources in the equations for the cooperative fields}\label{MinEMVa3}

\subsubsection{Axionic generalization of the Vlasov kinetic
equation}\label{MinEMVa31}

The relativistic collisionless kinetic equation based on the Vlasov approach is of the form
\begin{equation}
\frac{p^i}{m_{({\rm a})}} \left[\frac{\partial}{\partial x^i} {-} \Gamma^k_{il}p^l \frac{\partial}{\partial p^k}\right]f_{({\rm a})} {+}
\frac{\partial}{\partial p^k} \left[ {\cal F}^k_{({\rm a})}f_{({\rm a})}\right] {=} 0
\,,
\label{k3}
\end{equation}
where $f_{({\rm a})}$ is the distribution function of the particles of the sort ${({\rm
a})}$, $p^i$ is the momentum four-vector of the particle with the mass $m_{({\rm
a})}$. The four-vector ${\cal F}^k_{({\rm a})}$ plays a role of effective
force, which acts on the charged particle in the axion-active plasma.

\subsubsection{Macroscopic moments}\label{MinEMVa32}

In the Vlasov model the electric current four-vector $I^i$ is considered to consist in a linear combination of first
moments of the distribution functions
\begin{equation}
I^i = \sum_{({\rm a})} e_{({\rm a})} \int dP  f_{({\rm a})} \ p^i \,,
\label{k2}
\end{equation}
where $e_{({\rm a})}$ denotes the electric charge of the particle
of the sort ${({\rm a})}$; $dP= \sqrt{-g} \ d^4p$ is the invariant
integration volume in the momentum four-dimensional space. The
stress-energy tensor of the particles is given by the second
moment of the distribution function
\begin{equation}
{\cal T}^{({\rm m}) ik} = \sum_{({\rm a})} \int dP  f_{({\rm a})} p^i p^k \,. \label{01EineqMIN}
\end{equation}
Since the particle momentum four-vector is normalized ($g_{ik} p^i p^k {=} m^2_{({\rm
a})}$), the trace of the stress-energy tensor is
\begin{equation}
{\cal T}^{({\rm m})} = \sum_{({\rm a})}{\cal T}_{({\rm a})} = g_{ik} {\cal T}^{({\rm m})ik}  = \sum_{({\rm a})} m^2_{({\rm
a})} \int dP  f_{({\rm a})}  \,, \label{0145}
\end{equation}
i.e., it can be represented by the zero-order moment.
The pseudoscalar source ${\cal J}$ can also be considered as a zero-order moment of the distribution function
\begin{equation}
{\cal J} = \sum_{({\rm a})} \int dP  f_{({\rm a})} {\cal G}_{({\rm a})}   \label{j1}
\end{equation}
with the pseudoscalar quantity ${\cal G}_{({\rm a})}$, which  can, in principle, be decomposed with respect to the particle momentum four-vector $p^k$
as follows
\begin{equation}
{\cal G}_{({\rm a})} = \alpha_{({\rm a})} \phi + \beta_{({\rm a})}
p^k \nabla_k \phi + \gamma_{({\rm a})} p^k F^m_{\ k} \nabla_m \phi
+ \ldots \,. \label{j5}
\end{equation}
The coefficients in this decomposition are introduced phenomenologically, but below we consider the constraints for them coming from compatibility conditions.

\subsubsection{Reconstruction of the effective force using the compatibility conditions}\label{MinEMVa33}

The first compatibility condition (\ref{2axi}) with (\ref{k2}) and (\ref{k3}) yields
\begin{equation}
\nabla_k I^k = - \sum_{({\rm a})} e_{({\rm a})} m_{({\rm a})} \int dP \frac{\partial}{\partial p^k} \left[{\cal F}^k_{({\rm a})} f_{({\rm a})}\right] \equiv 0 \,, \label{0125}
\end{equation}
i.e., this condition is satisfied for arbitrary force-like term ${\cal F}^k_{({\rm
a})}$. The divergence of the stress-energy tensor yields
\begin{gather}
\nabla_k {\cal T}^{({\rm m}) ik} {=} \nabla_k \sum_{({\rm a})}
\int dP f_{({\rm a})} p^i p^k = - \sum_{({\rm a})} m_{({\rm a})}
\int dP p^i \frac{\partial}{\partial p^k} \left[{\cal F}^k_{({\rm
a})} f_{({\rm a})}\right] = \sum_{({\rm a})} m_{({\rm a})} \int dP
f_{({\rm a})} {\cal F}^i_{({\rm a})}, \label{check1}
\end{gather}
thus, according to (\ref{4axi}) and (\ref{k2}), the force-like
term ${\cal F}^i_{({\rm a})}$ has to satisfy the condition
\begin{gather}
\sum_{({\rm a})} \int dP f_{({\rm a})} \left[  m_{({\rm a})}{\cal
F}^i_{({\rm a})} {-} e_{({\rm a})} F^i_{\ m} p^m \right] = {\cal
J} \nabla^i \phi - \nabla_k T^{({\rm D})ik} \,. \label{check2}
\end{gather}
For the sake of simplicity we suppose below that the stress-energy
tensor $T^{({\rm D})ik}$ is divergence-free, i.e., $\nabla_k
T^{({\rm D})ik}{=}0$, but of course, it is easy to enlarge the
formalism discussed below for the case, when the energy-momentum
of the dark fluid is not a conserved quantity. Then according to
(\ref{check2}) the term ${\cal F}^k_{({\rm a})}$  splits into the
standard Lorentz force linear in the particle momentum
four-vector, and the force $\frak{R}^i_{({\rm a})}$ induced by the
axion field
\begin{equation}
m_{({\rm a})} {\cal F}^i_{({\rm a})} = e_{({\rm a})} F^i_{\ s} p^s
+ \frak{R}^i_{({\rm a})} \,. \label{check3}
\end{equation}
For the quantity $\frak{R}^i_{({\rm a})}$ we obtain the integral
equation
\begin{equation}
\sum_{({\rm a})} \int dP f_{({\rm a})} \left[ \frak{R}^i_{({\rm
a})} - {\cal G}_{({\rm a})} \nabla^i \phi \right] =0 \,.
\label{check37}
\end{equation}
There are three principal possibilities to resolve this integral equation. Let us consider them in more detail.

\vspace{3mm}

\noindent
{\it (i) There are no contact plasma-axion interactions}

\noindent The simplest version assumes that axions interact with
plasma through the electromagnetic field only, so that the
Lagrangian $L_{({\rm m})}$ does not contain the pseudoscalar field
$\phi$. This means that the source term ${\cal J}$ vanishes and
the force $\frak{R}^i_{({\rm a})}$ is absent. Thus the
compatibility conditions (\ref{check37}) are assumed to be
satisfied identically.

\vspace{3mm}

\noindent
{\it (ii) Microscopic force and particle rest-mass variation}

\noindent The second version relates to the reconstruction of a
{\it microscopic} force, which can include the particle momentum
four-vector $p^i$, the axion field $\phi$ and its derivatives.
Keeping in mind (\ref{check37}) such a force can be represented as
$\frak{R}^i_{({\rm a})} {=} {\cal G}_{({\rm a})} \nabla^i \phi$,
and clearly this force is not orthogonal to the particle momentum
four-vector $p^i$. This means that the particle rest mass is not
conserved. In order to illustrate such a model let us assume that
only the first term in (\ref{j5}) is nonvanishing, i.e., ${\cal
G}_{({\rm a})} {=} \alpha_{({\rm a})} \phi$, where $\alpha_{({\rm
a})}$ is constant. Then
\begin{equation}
\frac{1}{2} \frac{d}{d\tau} \left[p_i p^i \right] =
\frac{1}{m_{({\rm a})}}\frak{R}^i_{({\rm a})} p_i =  \frac{1}{2}
\frac{d}{d\tau} \left[\alpha_{({\rm a})} \phi^2 \right]\,,
\label{check004}
\end{equation}
and the square of the particle momentum four-vector
\begin{equation}
p_i p^i = m^2_{({\rm a})} + \alpha_{({\rm a})} \phi^2
\label{check005}
\end{equation}
depends on coordinates through the axion field $\phi$.

\vspace{3mm}

\noindent
{\it (iii) Ponderomotive force in plasma induced by axions}

\noindent The third variant to solve the integral equation
(\ref{check37}) is to reconstruct a {\it macroscopic} force, which
can contain some macroscopic moments of the distribution function;
we indicate the force of this type as the ponderomotive force in
analogy with the force appearing in the electrodynamics of
continuous media.  In our case the macroscopic force can be found
using the following ansatz. Let the force be orthogonal to the
particle four-momentum (thus providing the mass conservation), be
quadratic in $p^k$ and have the form
\begin{equation}
\frak{R}^i_{({\rm a})} = \left[\delta^i_k \ (p_s p^s) - p^i p_k
\right] \nu_{({\rm a})}  {\cal B}^k  \,, \quad \frak{R}^i_{({\rm
a})} p_i \equiv 0 \,, \label{check4}
\end{equation}
where $\nu_{({\rm a})}$ are constants introduced
phenomenologically, and ${\cal B}^k $ are unknown functions. Then
one obtains
\begin{equation}
\sum_{({\rm a})} \int dP f_{({\rm a})} \frak{R}^i_{({\rm a})} =
{\cal S}^i_k \ {\cal B}^k   \,, \label{check5}
\end{equation}
where
\begin{equation}
{\cal S}^i_k \equiv \left[\delta^i_k \ S_m^m - S^i_k \right] \,, \quad S^i_k  = \sum_{({\rm a})} \nu_{({\rm a})} {\cal T}^i_{k ({\rm a})}\,. \label{check55}
\end{equation}
Now, the condition (\ref{check2}) yields that the term ${\cal B}^k $ can be found using matrix equation
\begin{equation}
{\cal S}^i_k {\cal B}^k = {\cal J} \nabla^i \phi \,, \quad {\cal B}^k = \tilde{{\cal S}}^k_l  \ {\cal J} \nabla^l \phi \,, \label{check7}
\end{equation}
where $\tilde{{\cal S}}^i_k $ is the reciprocal matrix to the
matrix ${\cal S}^i_k $, i.e., $\tilde{{\cal S}}^i_k {\cal S}^k_l =
\delta^i_l$. Of course, we assume that the matrix ${\cal S}^i_k$
is nondegenerated, i.e., $\det {\cal S}^i_k \neq 0$. Finally, the
force, which we search for, has the form
\begin{equation}
\frak{R}^i_{({\rm a})} = \left[\delta^i_k (p_s p^s) - p^i p_k
\right] \nu_{({\rm a})} \tilde{{\cal S}}^k_l {\cal J} \nabla^l
\phi \,. \label{check8}
\end{equation}
In order to illustrate the described procedure let us consider the
following example. Let $\nu_{({\rm a})}= \nu$ (i.e., these
parameters do not depend on sort of particles $({\rm a})$), and
let the stress-energy tensor of the plasma particles be described
by the tensor for perfect fluid, i.e.,
\begin{equation}
{\cal T}^i_k = (W+P) U^i U_k - \delta^i_k P \,.
\label{ex1}
\end{equation}
Then the tensors  ${\cal S}^i_k$ and $\tilde{{\cal S}}^k_l$ are, respectively,
\begin{equation}
{\cal S}^i_k = \nu \left[\delta^i_k (W-2P) - (W+P) U^i U_k \right]\,,
\label{ex2}
\end{equation}
\begin{equation}
\tilde{{\cal S}}^k_l = \frac{1}{3\nu P(W-2P)} \left[3P \delta^k_l - (W+P) U^k U_l \right]
\,.
\label{ex21}
\end{equation}
Here $W$ is the energy-density scalar, $P$ is the Pascal pressure,
$U^i$ is the macroscopic velocity four-vector of the plasma system
as a whole. Clearly, the reciprocal tensor exists, when $P \neq 0$
and $W \neq 2P$. There is  one interesting subcase: when $W=5P$
one obtains from (\ref{ex2})
$$
{\cal S}^i_k = \frac{3\nu}{5} W {\cal L}^i_k \,, \quad \tilde{{\cal S}}^k_l = \frac{5}{3\nu W} {\cal L}^k_l \,,
$$
\begin{equation}
{\cal L}^i_k \equiv
\delta^i_k - 2 U^i U_k \,, \quad {\cal L}^i_k {\cal L}^k_l = \delta^i_l \,.
\label{ex3}
\end{equation}
Keeping in mind the standard equation of state $P=(\gamma-1)W$, we
see that this case relates to the value $\gamma=\frac{6}{5}$,
i.e., the adiabatic index $\gamma$ coincides with the upper
critical value obtained in the Lane-Emden theory of Newtonian
stars (see, e.g., \cite{Wbook}).

\subsubsection{Entropy balance equation}\label{MinEMVa34}

The entropy production of the kinetic system, the evolution of
which is guided by the kinetic equation (\ref{k3}) is connected
with the action of the force $\Re^i_{({\rm a})}$ ($k_{\rm (B)}$ is
the Boltzmann constant, $h=2\pi\hbar$):
\begin{gather}
\sigma = \nabla_i S^i = - k_{({\rm B})} \nabla_i \sum_{({\rm a})}
\int dP f_{({\rm a})} p^i \left[\log{(h^3f_{({\rm
a})})}{-}1\right] = k_{({\rm B})} \sum_{({\rm a})}m_{({\rm a})}
\int dP \log{(h^3f_{({\rm a})})} \frac{\partial}{\partial
p^i}\left[{\cal
F}^i_{({\rm a})}f_{({\rm a})} \right] ={}\nonumber\\
{}=  k_{({\rm B})} \sum_{({\rm a})} \int dP f_{({\rm a})}
\frac{\partial}{\partial p^i} \frak{R}^i_{({\rm a})} \,.
\label{F6}
\end{gather}
Clearly, the entropy production is absent when $\frak{R}^i_{({\rm
a})}$ is vanishing or does not depend on $p^k$, as it was in the
case of microscopic force reconstruction. When the ponderomotive
force exists, we find using (\ref{check8}) that the entropy
production scalar is nonvanishing and is proportional to the
pseudoscalar source-term ${\cal J}$:
\begin{equation}
\sigma = - 3 k_{({\rm B})} {\cal J} \tilde{{\cal S}}^l_k  \nabla_l \phi \sum_{({\rm a})} \nu_{({\rm a})} \int dP f_{({\rm a})} p^k
\,.
\label{F7}
\end{equation}
When ${\cal J}{=}0$, we see that $\sigma{=}0$, and the plasma is in the state of local equilibrium.

\subsubsection{On distribution functions of the equilibrium type}\label{MinEMVa35}

Equilibrium-type distribution function
\begin{equation}
f^{({\rm eq})}_{({\rm a})} = h^{-3} \exp\left\{\mu_{({\rm a})} - \xi^{({\rm a})}_k p^k \right\}
\label{Eq1}
\end{equation}
satisfies the kinetic equation (\ref{k3}) with the force
(\ref{check3}), (\ref{check4}), when
\begin{gather}
p^i \left[\frac{\partial}{\partial x^i} \mu_{({\rm a})}  {+}
e_{({\rm a})} F_{ik} \xi^{k {({\rm a})}} {-} 3 \nu_{({\rm
a})}{\cal B}_i \right]  - p^i p^k \left[ \nabla_i \xi^{({\rm
a})}_k {-} \xi^{({\rm a})}_i \nu_{({\rm a})}{\cal B}_k {+} g_{ik}
\xi^{s {({\rm a})}} \nu_{({\rm a})} {\cal B}_s \right] {=} 0 \,.
\label{Eq2}
\end{gather}
The momentum four-vector in this context is considered to be a
random variable, thus, one obtains two sets of conditions, which
are necessary the  equation (\ref{Eq2}) to be satisfied. The first
set of conditions
\begin{equation}
\frac{\partial}{\partial x^i} \mu_{({\rm a})} {=} {-} e_{({\rm a})} F_{ik} \xi^{k {({\rm a})}} {+} 3 \nu_{({\rm a})} {\cal B}_i \,,
\label{Eq3}
\end{equation}
contains four equations for one unknown function $\mu_{({\rm a})}$ (for the fixed index of the sort $({\rm a})$). The integrability conditions
$\frac{\partial^2}{\partial x^i \partial x^j} \mu_{({\rm a})} = \frac{\partial^2}{\partial x^j \partial x^i} \mu_{({\rm a})}$ are satisfied, when
\begin{gather}
e_{({\rm a})}\left[\frac{\partial}{\partial x^i} \pounds_{\xi}
A_{j} - \frac{\partial}{\partial x^j} \pounds_{\xi} A_{i} \right]
= - 3 \nu_{({\rm a})} \left[\frac{\partial}{\partial x^i} {\cal
B}_j - \frac{\partial}{\partial x^j}{\cal B}_i \right] \,.
\label{Eq31}
\end{gather}
Here the symbol $\pounds_{\xi}$ denotes the Lie derivative along the four-vector $\xi^{i ({\rm a})}$
\begin{equation}
\pounds_{\xi} A_{j} \equiv \xi^{l ({\rm a})} \frac{\partial}{\partial x^l} A_j + A_l \frac{\partial}{\partial x^j} \xi^{l ({\rm a})} \,.
\label{Eq9}
\end{equation}
The second term in (\ref{Eq2}), quadratic in the particle momentum
four-vector,  gives the condition
\begin{gather}
\pounds_{\xi} g_{ik} \equiv \nabla_i \xi^{({\rm a})}_k + \nabla_k
\xi^{({\rm a})}_i = \nu_{({\rm a})} \left[\xi^{({\rm a})}_i {\cal
B}_k {+} \xi^{({\rm a})}_k {\cal B}_i  {-} 2 g_{ik} \xi^{s {({\rm
a})}} {\cal B}_s \right]\,. \label{Eq4}
\end{gather}
When ${\cal B}_s {=} 0$ it is the well-known Killing equation.
When the right-hand side of the equation (\ref{Eq4}) is not
vanishing, we deal with a generalization of the Killing equation.
In a particular case, when  $\xi^{({\rm a})}_i {=} \xi_i$, and
$\nu_{({\rm a})}{=}\nu$ i.e., they do not depend on the sort of
particle, and ${\cal B}_i {=} {\cal B} \xi_i$, the equation
(\ref{Eq4}) reduces to the equation
\begin{equation}
\pounds_{\xi} g_{ik} = -2 \nu {\cal B} \left[g_{ik} \xi^s \xi_s  - \xi_i \xi_k \right]\,,
\label{Eq401}
\end{equation}
discussed in the works \cite{tide2,tide5} in the context of
the antifriction force and accelerated expansion of the Universe.
The equilibrium state in plasma exists if the equation (\ref{Eq4})
admits the existence of the time-like four-vectors $\xi^{({\rm
a})}_i$, and if they coincide for all sorts of particles, i.e.,
$\xi^{({\rm a})}_i = \xi_i$. The solution of (\ref{Eq31}) is
\begin{equation}
e_{({\rm a})}\pounds_{\xi} A_{j} = - 3 \nu_{({\rm a})} {\cal B}_j + \frac{\partial}{\partial x^j} \Psi_{({\rm a})} \,,
\label{Eq91}
\end{equation}
where $\Psi_{({\rm a})}$ is arbitrary scalar. Let us consider two different cases.

\vspace{3mm}

\noindent
{\it (i) General case $A_i \neq 0$}

\noindent
When $A_i \neq 0$, the scalar $\Psi_{({\rm a})}$ can be eliminated using the gauge transformation of the potential four-vector
\begin{equation}
A_{j} \to {\tilde A}_j + \nabla_j {\tilde \Psi} \,, \quad  \Psi_{({\rm a})} = e_{({\rm a})} \xi^{l ({\rm a})} \frac{\partial}{\partial x^l} {\tilde \Psi}\,,
\label{Eq92}
\end{equation}
thus $\Psi_{({\rm a})}$ can be put equal to zero. When (\ref{Eq91}) is satisfied, the scalar function $\mu_{({\rm a})}$ can be found from the equation
\begin{gather}
\frac{\partial}{\partial x^i} \mu_{({\rm a})} = - e_{({\rm a})}
\left[F_{ik} \xi^{k {({\rm a})}} + \pounds_{\xi} A_{i} \right] = -
e_{({\rm a})} \frac{\partial}{\partial x^i} \left[\xi^{k {({\rm
a})}} A_{k} \right] \,, \label{Eq37}
\end{gather}
yielding
\begin{equation}
\mu_{({\rm a})} = {\tilde \mu}^{(0)}_{({\rm a})} - e_{({\rm a})}
\xi_k^{{({\rm a})}} A^{k}
 \,,
\label{Eq38}
\end{equation}
where ${\tilde \mu}^{(0)}_{({\rm a})}$ is a constant. Thus, when,
first, the four-vector $\xi^i{=}\xi^i_{({\rm a})}$ is timelike and
satisfies the equation (\ref{Eq4}), second, the equation
(\ref{Eq91}) is satisfied, then, using the standard definitions of
the chemical potentials $\mu^{(0)}_{({\rm a})}$, of the
macroscopic velocity four-vector $U^i$ and of the temperature $T$
\begin{equation}
{\tilde \mu}^{(0)}_{({\rm a})} = \frac{\mu^{(0)}_{({\rm a})}}{k_{({\rm B})}T} \,, \quad \xi^i = \frac{U^i}{k_{({\rm B})}T}
\,,
\label{Eq41}
\end{equation}
one can rewrite the equilibrium type function (\ref{Eq1}) in the well-known form
\begin{equation}
h^3 f^{({\rm eq})}_{({\rm a})} {=}
\exp\left\{\frac{\mu^{(0)}_{({\rm a})} {-} U_i \left[p^i {+} e_{({\rm a})} A^{i} \right]}{k_{({\rm B})}T} \right\}
\delta\left[\sqrt{p^ip_i}{-}m_{({\rm a})}\right].
\label{Eq101}
\end{equation}
Let us mention that in this case the force-like term (\ref{check4}), quadratic in the particle four-momentum
\begin{equation}
\frak{R}^i_{({\rm a})} = - \frac{1}{3}\left[g^{ik} \ (p_s p^s) -
p^i p^k \right] e_{({\rm a})} \pounds_{\xi} A_{k} \label{check41}
\end{equation}
is proportional to the Lie derivative of the electromagnetic potential four-vector.

\vspace{3mm}

\noindent
{\it (ii) Special case $A_i {=} 0$}

\noindent
When the macroscopic electromagnetic field in the electroneutral plasma vanishes, and thus $A_i=0$, the equation (\ref{Eq91}) is satisfied, if the four-vector ${\cal B}_j$ is the gradient four-vector
\begin{equation}
{\cal B}_j =  \frac{1}{3 \nu_{({\rm a})} }\frac{\partial}{\partial x^j} \Psi_{({\rm a})} \,, \quad \mu_{({\rm a})} =  {\tilde \mu}^{(0)}_{({\rm a})} - \Psi_{({\rm a})} \,.
\label{Eq99}
\end{equation}
As for the equation (\ref{Eq4}), it now is of the form
\begin{equation}
\pounds_{\xi} g_{ik} = \frac{1}{3}\left[\xi^{({\rm a})}_i \nabla_k
\Psi_{({\rm a})} {+} \xi^{({\rm a})}_k \nabla_i \Psi_{({\rm a})}
{-} 2 g_{ik} \xi^{s {({\rm a})}} \nabla_s \Psi_{({\rm a})}
\right]. \label{Eq444}
\end{equation}
In other words, the contact interactions between axions and plasma particles can support an equilibrium state in the system, when the coupling parameters satisfy some specific conditions.

\section{Nonminimal extension of the Einstein-Maxwell-Vlasov-axion - plasma
model}\label{NMEMVa}

\subsection{Nonminimal extension of the Lagrangian}\label{NMEMVa1}

We consider now the total action functional as a sum of minimal and
nonminimal contributions
\begin{equation}
S = S_{({\rm M})} + S_{({\rm NM})} \,, \label{plus}
\end{equation}
where $S_{({\rm M})}$ is given by (\ref{R1}) and
\begin{gather}
S_{({\rm NM})} = \int d^4 x \sqrt{{-}g} \left\{ \frac{1}{4} {\cal
R}^{ikmn} F_{ik}F_{mn} + \frac{1}{4} {\chi}^{ikmn}_{({\rm A})}
\phi \ F_{ik} \Fst_{mn} -\frac{1}{2}\Psi^2_0\left[\Re^{mn}_{({\rm
A})} \nabla_m \phi \nabla_n \phi - \eta_{({\rm A})} R  \phi^2
\right]\right\} \,. \label{actnm}
\end{gather}
The quantity ${\cal R}^{ikmn}$ is a nonminimal three-parameter
susceptibility tensor \cite{BL05}, which has a form
\begin{equation}
{\cal R}^{ikmn} =  q_1 R g^{ikmn} + q_2 \Re^{ikmn} + q_3 R^{ikmn}
\,, \label{sus1}
\end{equation}
where the following auxiliary tensors are introduced
\begin{gather}
g^{ikmn} \equiv \frac{1}{2}(g^{im}g^{kn} {-} g^{in}g^{km}) \,,
\label{rrr}\\
\Re^{ikmn} \equiv \frac{1}{2} (R^{im}g^{kn} {-} R^{in}g^{km} {+}
R^{kn}g^{im} {-} R^{km}g^{in}) \,. \label{rrrr}
\end{gather}
The constants $q_1$, $q_2$ and $q_3$ are nonminimal parameters
describing the linear coupling of the Maxwell tensor $F_{mn}$ with
curvature \cite{BL05}. The quantity ${\chi}^{ikmn}_{({\rm A})}$,
the nonminimal susceptibility tensor describing the linear
coupling of the dual tensor $\Fst_{mn}$ with curvature, can be
initially modeled as
\begin{equation}
{\chi}^{ikmn}_{({\rm A})} = Q_1 R g^{ikmn} {+} Q_2\Re^{ikmn}
{+} Q_3 R^{ikmn} \,, \label{sus2}
\end{equation}
using the direct analogy with the decomposition (\ref{sus1}) and
introducing three phenomenological parameters $Q_1$, $Q_2$ and
$Q_3$. As in the previous case, the combination $\phi \Fst_{mn}$ gives
the tensor quantity, thus ${\chi}^{ikmn}_{({\rm A})}$ is also a
pure tensor. However, taking into account the relation
(\ref{5axi}) we conclude that the term ${\Re}^{ikmn} \phi F_{ik}
\Fst_{mn}$ is proportional to $R{g}^{ikmn} \phi \,F_{ik}
\Fst_{mn}$, and we have only two independent coupling constants
among three parameters $Q_1$, $Q_2$ and $Q_3$. In the next
subsection we discuss this problem in detail and motivate our
ansatz that $Q_2=-Q_3$.

The tensors ${\cal R}^{ikmn}$ and  ${\chi}^{ikmn}_{({\rm A})}$,
defined by (\ref{sus1}) and (\ref{sus2}), are skew-symmetric with
respect to transposition of the indices $i$ and $k$, as well as
$m$ and $n$. In addition the following relations take place
\begin{equation}
{\cal R}^{ikmn}= {\cal R}^{mnik} \,, \quad {\chi}^{ikmn}_{({\rm
A})} = {\chi}^{mnik}_{({\rm A})}\,, \label{sus8}
\end{equation}
which guarantee that the model under consideration does not
contain the solutions of the skewon type \cite{HehlObukhov}.
The symmetric tensor
\begin{equation}
\Re^{mn}_{({\rm A})} \equiv \frac{1}{2} \eta_1
\left(F^{ml}R^{n}_{\ l} + F^{nl}R^{m}_{\ l} \right) + \eta_2 R
g^{mn} + \eta_3 R^{mn} \label{sus3}
\end{equation}
describes a nonminimal susceptibility for the pseudoscalar field
in analogy with the Higgs fields \cite{BDZ1}, but in this case the
tensor $\Re^{mn}_{({\rm A})}$ contains an additional term linear
in the Maxwell tensor. This term describes effects analogous to
the so-called derivative coupling in the nonminimal scalar field
theory \cite{derc1,derc2}. As for the tenth coupling constant
$\eta_{({\rm A})}$, it is a direct analog of the well-known
coupling constant $\xi$ in the nonminimal scalar field theory
(see, e.g., \cite{FaraR} for review and references).

\subsection{One-parameter and two-parameters families of the models for nonminimal susceptibility tensor}\label{NMEMVa2}

The pseudoscalar ${\chi}^{ikmn}_{({\rm A})} F_{ik} \Fst_{mn}$ with
the nonminimal susceptibility tensor (\ref{sus2}) can be reduced
to the pseudoscalar
\begin{gather}
{\chi}^{ikmn}_{({\rm A})} F_{ik} \Fst_{mn} =
\left[\left(Q_1{+}\frac{1}{2}Q_2 \right) R g^{ikmn} {+} Q_3
R^{ikmn} \right] F_{ik} \Fst_{mn}  \label{AppB2}
\end{gather}
due to the relation (\ref{5axi}) proved in the Appendix A.
This means that, when we make the replacements
\begin{gather}
Q_1 = \overline{Q}_1 + {\cal Q} \,, \quad  Q_2 =\overline{Q}_2 -
2{\cal Q} \,,\quad Q_3=\overline{Q}_3 \label{gauge1}
\end{gather}
with arbitrary constant ${\cal Q}$, the invariant $
\frac{1}{4}{\chi}^{ikmn}_{({\rm A})} \phi F_{ik} \Fst_{mn}$ keeps
the form and the master equations obtained by the variation of the
corresponding term in the action functional (\ref{actnm}) remain
unchanged. In this sense we can consider the relations
(\ref{gauge1}) as an analog of gauge transformations. This fact
gives us an argument to focus on the problem of an appropriate
choice of the constants $Q_1$, $Q_2$, $Q_3$, which allows us to
simplify the model.

\subsubsection{Symmetry with respect to the left and right dualizations}

Let us assume that the nonminimal susceptibility tensor satisfies the following symmetry condition
\begin{equation}
{}^{*}{\chi}^{ikmn}_{({\rm A})} = {\chi}^{*ikmn}_{({\rm A})}  \,,
\label{sus27}
\end{equation}
which is equivalent to
\begin{equation}
{}^{*}{\chi}^{*ikmn}_{({\rm A})} = - {\chi}^{ikmn}_{({\rm A})}  \,.
\label{sus28}
\end{equation}
Taking into account the following relations for the double-dual quantities
\begin{gather}
{}^{*}{g}^{*ikmn} = - g^{ikmn}  \,,\label{sus29}\\
{}^{*}{\Re}^{*ikmn} =  {\Re}^{ikmn} - R g^{ikmn}  \,,
\label{sus30}\\
{}^{*}{R}^{*ikmn} = -R^{ikmn}+ 2 \Re^{ikmn}- Rg^{ikmn}  \,,
\label{sus31}
\end{gather}
we can conclude that the symmetry condition (\ref{sus27}) takes
place for arbitrary curvature tensor, when $Q_2{+}Q_3=0$. This
condition has been implicitly used in the work \cite{BaNi}. The
pseudoscalar (\ref{AppB2}) takes now the form
\begin{gather}
{\chi}^{ikmn}_{({\rm A})} F_{ik} \Fst_{mn} =
\left[\left(Q_1{-}\frac{1}{2}Q_3 \right) R g^{ikmn} {+} Q_3
R^{ikmn} \right] F_{ik} \Fst_{mn} \,. \label{AppB21}
\end{gather}
Using the Weyl tensor ${\cal C}^{ikmn}$ and the standard decomposition of the Riemann tensor, we can also represent the susceptibility tensor as follows
\begin{gather}
{\chi}^{ikmn}_{({\rm A})} = Q_3 {\cal C}^{ikmn} +
(Q_2+Q_3)\Re^{ikmn} +\left( Q_1-\frac{1}{3}Q_3\right) R g^{ikmn}
\,. \label{sus34}
\end{gather}
The condition $Q_2{+}Q_3=0$ excludes the term $\Re^{ikmn}$ from this
decomposition, since only the Weyl tensor ${\cal C}^{ikmn}$ and
the tensor $g^{ikmn}$ possess symmetry with respect to left and
right dualizations.

Since the constraint $\overline{Q}_2{+}\overline{Q}_3=0$ will always
be satisfied by the replacements (\ref{gauge1}) with the suitable
choice of the fitting parameter ${\cal Q}= -\frac{1}{2}(Q_2+Q_3)$,
without loss of generality, we can assume that the nonminimal
susceptibility tensor ${\chi}^{ikmn}_{({\rm A})}$ obeys the
symmetry condition (\ref{sus27}). Thus our model is effectively
two-parameters and one parameter, say, $Q_2$, is the hidden one.
Besides, we can impose an additional constraint on the constants
$Q_1$, $Q_2$, $Q_3$, and therefore reduce our general model to a
certain one-parameter submodel. Below we consider several examples
of such submodels.

\subsubsection{One-parameter submodels}

Let us remind that the tensor ${\cal R}^{im}$
\begin{equation}
{\cal R}^{im} \equiv g_{kn} {\cal R}^{ikmn} = \frac{1}{2}R
g^{im}(3q_1+q_2) + R^{im}(q_2+q_3) \,, \label{susca1}
\end{equation}
vanishes in a generic (curved) spacetime, when the nonminimal
coupling parameters are linked by two relations $3q_1{+}q_2=0$ and
$q_2{+}q_3=0$. Analogously, the scalar ${\cal R}$
\begin{equation}
{\cal R} \equiv g_{im}g_{kn} {\cal R}^{ikmn} = R (6q_1+3q_2+q_3)
\,, \label{susca}
\end{equation}
takes zero value in the curved spacetime, when
$6q_1{+}3q_2{+}q_3=0$. Similarly, we obtain that
\begin{equation}
{\chi}^{im}_{({\rm A})} \equiv g_{kn} {\chi}^{ikmn}_{({\rm A})}
{=} \frac{1}{2}R g^{im}(3Q_1+Q_2) + R^{im}(Q_2+Q_3) \,,
\label{susca11}
\end{equation}
vanishes, when $3Q_1+Q_2=0$ and $Q_2+Q_3=0$, and
\begin{equation}
{\chi}_{({\rm A})} \equiv g_{im}g_{kn} {\chi}^{ikmn}_{({\rm A})} =
R (6Q_1+3Q_2+Q_3) \,, \label{susca13}
\end{equation}
takes zero value, when $6Q_1+3Q_2+Q_3=0$. Clearly, the condition
$6Q_1+3Q_2+Q_3 =6\overline{Q}_1+3\overline{Q}_2+\overline{Q}_3=0$
is invariant with respect to the transformation (\ref{gauge1}),
other conditions require special choices of the parameter ${\cal
Q}$.

\paragraph{The model with vanishing scalar ${\chi}_{({\rm
A})}$.} When $6Q_1{+}3Q_2{+}Q_3=0$ and thus ${\chi}_{({\rm
A})}=0$, one obtains that the pseudoscalar (\ref{AppB2}) turns
into
\begin{equation}
{\chi}^{ikmn}_{({\rm A})} F_{ik} \Fst_{mn} = Q_3 \left[R^{ikmn} -
\frac{1}{6}R g^{ikmn}\right] F_{ik} \Fst_{mn} \,. \label{AppB29}
\end{equation}
Since the term $6Q_1{+}3Q_2{+}Q_3$ is invariant with respect to
the transformation (\ref{gauge1}), this model contains only two
arbitrary parameters, say, $Q_2$ and $Q_3$, but now $Q_2$ does not
enter the invariant (\ref{AppB29}), i.e., it becomes effectively
hidden.

\paragraph{Weyl-type relation.} A model, in which $3Q_1{+}Q_2=0$ and $Q_2{+}Q_3=0$, appears when
we suggest that ${\chi}_{({\rm A})}^{mn}=0$, and thus the
susceptibility tensor is proportional to the Weyl tensor
\begin{equation}
{\chi}_{({\rm A})}^{ikmn} = Q_3 {\cal C}^{ikmn} \,. \label{R51}
\end{equation}
These conditions are equivalent to $6Q_1{+}3Q_2{+}Q_3=0$ and
$Q_2{+}Q_3=0$, where the first one is invariant with respect to
(\ref{gauge1}), while the second constraint just fixes the fitting
parameter (see above). Therefore we can conclude that the previous
condition with vanishing scalar ${\chi}_{({\rm A})}$ comes to this
one by the suitable choice of ${\cal Q}$, and the pseudoscalar
(\ref{AppB2}) takes the same form (\ref{AppB29}).

\paragraph{Gauss-Bonnet-type relation.} The model, for which the susceptibility tensor is proportional to
the double-dual Riemann tensor
\begin{equation}
{\chi}^{ikmn}_{({\rm A})} = - Q_3  \ ^{*}R^{*\,ikmn} \,,
\label{AppB39}
\end{equation}
is indicated in \cite{BL05} as the Gauss-Bonnet-type model; such
susceptibility tensor is divergence-free. In our context, this
proportionality yields that the nonminimal parameters are coupled
by two relations $Q_1{-}Q_3{=}0$ and $2Q_1{+}Q_2{=}0$, and the
pseudoscalar (\ref{AppB2}) is of the form
\begin{equation}
{\chi}^{ikmn}_{({\rm A})} F_{ik} \Fst_{mn} = Q_3 R^{ikmn} F_{ik}
\Fst_{mn}\,. \label{R5}
\end{equation}
The condition $\overline{Q}_1-\overline{Q}_3=Q_1-Q_3 +{\cal Q}=0$
can be satisfied by the choice ${\cal Q}= Q_3-Q_1$, i.e., the
value of the fitting parameter $\cal Q$ is fixed. The condition
$2Q_1+Q_2 = 2\overline{Q}_1+\overline{Q}_2=0$ is invariant with
respect to the transformations (\ref{gauge1}), thus, this
requirement makes the model one-parameter. Let us mention, that
the choice ${\cal Q}=Q_1$, provides the relation
$\overline{Q}_1=\overline{Q}_2=0$ to be valid, thus we deal with
the model related to ${\chi}^{ikmn}_{({\rm A})} = Q_3 R^{ikmn}$.

\subsection{Nonminimal electrodynamic equations}\label{EDeqs}

Electrodynamic equations, which correspond to the action
functional (\ref{plus}) with (\ref{R1}) and (\ref{actnm}), are
linear and have the standard form
\begin{equation}
\nabla_k  H^{ik} = - I^i \,. \label{max2}
\end{equation}
The excitation tensor $H^{ik}$ and the Maxwell tensor $F_{mn}$ are
linked by the linear constitutive law (here we assume that
$Q_2=-Q_3$)
\begin{gather}
H^{ik} \equiv {\cal H}^{ik} + F^{ik} + {\cal R}^{ikmn} F_{mn} +
\left[\phi \left( \Fst^{ik} + {\chi}^{ikmn}_{({\rm A})} \Fst_{mn}
\right)\right]. \label{inducnm}
\end{gather}
The first term ${\cal H}^{ik}$ given by
\begin{equation}
{\cal H}^{ik} \equiv -\frac{1}{2}\eta_1 \Psi^2_0 \left[ \left(R^{km} \nabla^i
\phi - R^{im} \nabla^k \phi \right) \nabla_m \phi \right] \,,
\label{current}
\end{equation}
does not contain the Maxwell tensor and thus presents the so-called spontaneous polarization-magnetization tensor;
in our case this quantity relates to the nonminimal polarization-magnetization of the axionically active medium, since it is
linear in the Ricci tensor on the one hand, and in the four-gradient of the pseudoscalar (axion) field, on the other hand.
The third term on the right-hand side, which is linear in the Maxwell tensor, is the curvature induced
polarization-magnetization, appeared in the nonminimally
extended pure Einstein-Maxwell model \cite{BL05}. The contribution
detailed in square brackets describes the axion-photon coupling, the
terms $\phi$ and $\Fst^{ik}$ enter the equations in the
multiplicative form only. The nonminimal axion contribution $ \phi
{\chi}^{ikmn}_{({\rm A})} \Fst_{mn}$ is a new term in comparison
with the minimal model.
The four-vector of the electric current has the standard Vlasov form (\ref{k2}) as in the minimal model; again, it  satisfies
the conservation law $\nabla_i I^i {=} 0$.

\subsection{Nonminimal equation for the pseudoscalar (axion) field}\label{Aeqs}

Nonminimally extended master equation for the pseudoscalar $\phi$
takes the form
\begin{gather}
\nabla_m \left[ \left( g^{mn} {+} \Re^{mn}_{({\rm A})} \right)
\nabla_n \phi \right] {+} \left[m^2_{({\rm A })} {+}
V^{\prime}(\phi^2){+} \eta_{({\rm A})} R \right] \phi = {}\nonumber \\
{}= - \frac{1}{\Psi^2_0}\left[\sum_{({\rm a})} \int dP f_{({\rm
a})} {\cal G}_{({\rm a})} {+} \frac{1}{4} \Fst_{mn}\left(
F^{mn}{+} {\chi}^{ikmn}_{({\rm A})} F_{ik} \right) \right],
\label{eqaxi1}
\end{gather}
where $\Re^{mn}_{({\rm A})}$ and ${\chi}^{ikmn}_{({\rm A})}$ are
given by (\ref{sus3}) and (\ref{sus2}), respectively. This
equation is a nonminimal generalization of the master equation (\ref{induc3}).

\subsection{Nonminimal generalization of the equations for the gravitational field}\label{Geqs}

Variation of the action functional (\ref{plus}) with
(\ref{R1}) and (\ref{actnm}) with respect to $g^{ik}$ gives
the nonminimally extended equations for the gravitational field
\begin{gather}
R_{ik}{-}\frac{1}{2}Rg_{ik} - \Lambda g_{ik} = \kappa \left[
T^{({\rm m})}_{ik} {+} T^{({\rm EM})}_{ik}  {+} T^{({\rm A})}_{ik}
{+} T^{({\rm NMEM})}_{ik} {+} {\cal T}^{({\rm NMA})}_{ik} \right]
\,. \label{Eineq}
\end{gather}
with the tensors $T^{({\rm EM})}_{ik}$ and $T^{({\rm A})}_{ik}$
given by (\ref{1TEM}) and (\ref{TAX}), respectively. The
nonminimal extension of the stress-energy tensor contains two
contributions: first, $T^{({\rm NMEM})}_{ik}$ describing pure
nonminimal electromagnetic part (see, e.g., \cite{BL05} for
details), second, the nonminimal axion part ${\cal T}^{({\rm
NMA})}_{ik}$. These tensors can be specified as follows:
\begin{equation}
T^{({\rm NMEM})}_{ik} = q_1 T^{(1)}_{ik} + q_2 T^{(2)}_{ik}+ q_3
T^{(3)}_{ik} \,, \label{decompEM}
\end{equation}
\begin{gather}
{\cal T}^{({\rm NMA})}_{ik} = \left(Q_1-\frac{1}{2}Q_3
\right){\cal T}^{(1)}_{ik} + Q_3 {\cal T}^{(3)}_{ik} + \Psi^2_0
\left[\eta_1 {\cal T}^{(4)}_{ik} + \eta_2 {\cal T}^{(5)}_{ik} +
\eta_3 {\cal T}^{(6)}_{ik} + \eta_{({\rm A})} {\cal T}^{(7)}_{ik}
\right]\,, \label{decompAX}
\end{gather}
where we put $Q_2=-Q_3$. Nine nonminimal contributions to the
stress-energy tensor can be divided into two groups. The first
group
\begin{gather}
T^{(1)}_{ik} = \frac{1}{2} \left[ \nabla_{i} \nabla_{k} - g_{ik}
\nabla^l \nabla_l \right] \left[F_{mn}F^{mn} \right] - R
F_{im}F_{k}^{ \ m}  - \frac{1}{2} F_{mn}F^{mn}
\left(R_{ik}{-}\frac{1}{2}Rg_{ik}\right)
 \,, \label{T1}
\end{gather}
\begin{gather}
T^{(2)}_{ik} = -\frac{1}{2}g_{ik} \left[\nabla_{m}
\nabla_{l}\left(F^{mn}F^{l}_{\ n} \right) - R_{lm} F^{mn} F^{l}_{\
n} \right]- F^{ln} \left(R_{il}F_{kn} + R_{kl}F_{in} \right)-
\frac{1}{2}
\nabla^m \nabla_m \left(F_{in} F_{k}^{ \ n}\right)+{}\nonumber\\
{}+\frac{1}{2}\nabla_l \left[ \nabla_i \left( F_{kn}F^{ln} \right)
{+} \nabla_k \left(F_{in}F^{ln} \right) \right] {-} R^{mn} F_{im}
F_{kn} \,, \label{T2}
\end{gather}
\begin{gather}
T^{(3)}_{ik} {=} \frac{1}{4}g_{ik} R^{mnls}F_{mn}F_{ls}{-}
\frac{3}{4} F^{ls} \left(F_{i}^{\ n} R_{knls} {+} F_{k}^{\
n}R_{inls} \right) - \frac{1}{2} \nabla_{m} \nabla_{n} \left[
F_{i}^{ \ n}F_{k}^{ \ m} + F_{k}^{ \ n} F_{i}^{ \ m} \right] \,,
\label{T3}
\end{gather}
does not contain pseudoscalar field. Other two terms
\begin{gather}
{\cal T}^{(1)}_{ik}  \equiv  \frac{1}{2} \left[ \nabla_{i}
\nabla_{k} {-} g_{ik} \nabla^l \nabla_l \right] \left[\phi
\Fst_{mn}F^{mn} \right] - \frac{1}{2} R_{ik}\phi \Fst_{mn}F^{mn}
\,, \label{calT1}
\end{gather}
\begin{gather}
{\cal T}^{(3)}_{ik}  \equiv {-}\frac{1}{2} \nabla_{m} \nabla_{n}
\left[ \phi \left(\Fst_i^{\ n} {F_k}^{m} {+} \Fst_k^{\ n}
{F_i}^{m} \right)\right]+ \frac{1}{4} \phi \Fst^{mn} \left(F_{il}
R^{l}_{ \ kmn} {+} F_{kl} R^{l}_{\ imn} \right) \,, \label{calT3}
\end{gather}
are linear in the pseudoscalar field $\phi$, and last four terms
\begin{gather}
{\cal T}^{(4)}_{ik} \equiv  \frac{1}{2}g_{ik} \left(R^l_n {-}
\nabla^l \nabla_n \right) \left( F^{nm} \nabla_m \phi \nabla_l
\phi \right) +\frac{1}{2} R^l_n \nabla_l \phi \left(F_i^{\ n}
\nabla_k \phi {+} F_k^{\ n} \nabla_i \phi \right)
+\frac{1}{4}\nabla^l \nabla_l \left[ \nabla_m \phi \left(
 F^m_{\ \ i} \nabla_k \phi {+}  F^m_{\ \ k} \nabla_i \phi
\right)\right]+{}\nonumber \\
{}+ \frac{1}{4}\nabla^l \left[ \nabla_i \left(F_k^{\ m}\nabla_m
\phi \nabla_l \phi \right) {+} \nabla_k \left(F_i^{\ m}\nabla_m
\phi \nabla_l \phi \right) \right] +\frac{1}{4}\nabla_m \left[
\nabla_i \left( F^{mn}\nabla_k \phi \nabla_n \phi  \right) {+}
\nabla_k \left( F^{mn}\nabla_i \phi
\nabla_n \phi  \right) \right] +{}\nonumber \\
{}+\frac{1}{2} F^{mn}   \left(R_{in}\nabla_k \phi {+}
R_{kn}\nabla_i \phi \right)\nabla_m \phi + \frac{1}{2}\left(R^m_i
F^n_{\ k} {+} R^m_k F^n_{\ i} \right) \nabla_m \phi \nabla_n \phi
\,, \label{calT4}
\end{gather}
\begin{gather}
{\cal T}^{(5)}_{ik}=  R \nabla_i \phi \nabla_k \phi + \left(g_{ik}
\nabla_n \nabla^n - \nabla_i \nabla_k \right)\left[\nabla_m \phi
\nabla^m \phi \right] +\nabla_m \phi \nabla^m \phi
\left(R_{ik}{-}\frac{1}{2}Rg_{ik}\right) \,,\label{calT5}
\end{gather}
\begin{gather}
{\cal T}^{(6)}_{ik} = \nabla_m \phi \left[R_i^m \nabla_k \phi +
R_k^m \nabla_i \phi \right] +\frac{1}{2}g_{ik}\left( \nabla_m
\nabla_n -R_{mn}\right)\left[ \nabla^m \phi \nabla^n \phi \right]
- \nabla^m \left[ \nabla_m \phi\; \nabla_i \nabla_k \phi \right]
\,.\label{calT6}
\end{gather}
\begin{equation}\label{calT7}
{\cal T}^{(7)}_{ik}= \left(\nabla_i \nabla_k - g_{ik} \nabla_m
\nabla^m \right){\phi}^2 - \left(R_{ik}{-}\frac{1}{2}Rg_{ik}\right) \phi^2\,,
\end{equation}
are quadratic either in the four-gradient $\nabla_k \phi$, or in the axion field itself.
In order to write these nine nonminimal contributions in an
appropriate way we have used the Bianchi identities
\begin{equation}
\nabla_i R_{klmn} + \nabla_l R_{ikmn} + \nabla_k R_{limn} = 0 \,,
\label{bianchi1}
\end{equation}
the properties of the Riemann tensor
\begin{equation}
R_{klmn} + R_{mkln} + R_{lmkn} = 0 \,, \label{bianchi2}
\end{equation}
as well as the rules for the commutation of covariant derivatives
\begin{equation}
(\nabla_l \nabla_k - \nabla_k \nabla_l) {\cal A}^i = {\cal A}^m
R^i_{\ mlk} \,. \label{nana}
\end{equation}

\subsection{Compatibility conditions for the nonminimal master equations}\label{NMEMVa6}

The Bianchi identity requires, as usual, that the total stress-energy tensor is divergence-free
\begin{equation}
\nabla^k \left[ T^{({\rm m})}_{ik}{+}
T^{({\rm EM})}_{ik}  {+} T^{({\rm A})}_{ik} {+} T^{({\rm
NMEM})}_{ik} {+} {\cal T}^{({\rm NMA})}_{ik} \right]  = 0
\,. \label{EinBI}
\end{equation}
In the Appendix B we show explicitly that this relation with
nonminimal tensors $T^{({\rm NMEM})}_{ik}$ and ${\cal T}^{({\rm
NMA})}_{ik}$, given by (\ref{decompEM})-(\ref{calT7}), yields
formally the same equation, as in the minimal model (see
(\ref{4axi})). Thus, the requirement (\ref{check37}) for the force
four-vector should be valid. Only one new detail appears: the
pseudoscalar ${\cal G}_{({\rm a})}$, which is given by (\ref{j5})
in the minimal case, can be extended as follows:
\begin{gather}
{\cal G}_{({\rm a})} \to {\cal G}_{({\rm a})} + \lambda_{({\rm
a})} \phi R + \mu_{({\rm a})} R \ p^k \nabla_k \phi + \nu_{({\rm
a})} R^m_n p^n \nabla_m \phi + \omega_{({\rm a})} R^i_{\ kmn} p^k
F^{mn} \nabla_i \phi +  \ldots \,, \label{j51}
\end{gather}
where the nonminimal part contains various convolutions of the
Riemann tensor with the particle momentum four-vector, the Maxwell
tensor, four-gradient of the axion field, etc.

\section{Conclusions}\label{Concl}

We formulated the nonminimal Einstein-Maxwell-Vlasov-axion model,
i.e., obtained the self-consistent system of master equations,
which describes the evolution of nonminimally coupled
gravitational (see Subsection~\ref{Geqs}), electromagnetic (see
Subsection~\ref{EDeqs}) and pseudoscalar (see
Subsection~\ref{Aeqs}) fields in the multi-component relativistic
plasma. We followed the combined approach: we used the Lagrange
formalism to derive the nonminimally extended field equations, and
the formalism of relativistic kinetic theory to link the
pseudoscalar, vectorial and tensorial sources in the right-hand
sides of the field equations with the macroscopic moments of the
plasma distribution function. Then we checked directly the
compatibility conditions to verify the self-consistency of this
combined approach. We prepared the model for the next step: for
cosmological and astrophysical applications.

In this work we follow the Vlasov concept of the cooperative
fields. First of all, the gravitational field is considered to be
the cooperative one: on the one hand, it governs the evolution of
the electromagnetic and pseudoscalar (axion) fields and the plasma
particle dynamics; on the other hand, these fields and plasma
particles form the corresponding sources for the gravity field
evolution. Second, we consider the cooperative electromagnetic
field generated in plasma as a macroscopic field averaged over the
statistical ensemble; the Lorentz force guiding the plasma
particle contains this macroscopic electromagnetic field, and the
electric current in the electrodynamic equations includes the
first-order macroscopic moment of the distribution function.
Third, evolution of the cooperative pseudoscalar (axion) field, on
the one hand, is regulated by the cooperative gravitational and
electromagnetic fields and by the pseudoscalar source induced in
plasma; on the other hand, this axion field contributes to the
total stress-energy tensor, the source for the gravitational
field, forms specific current-like source in the electrodynamic
equations, and acts on the plasma particles via the force appeared
in the relativistic kinetic equation.

Only one element of the model is not yet fixed explicitly: the
density ${\cal G}_{({\rm a})}$ of the pseudoscalar source ${\cal
J}$ appeared in the master equation for the axion field
(\ref{eqaxi1}). Its structure is assumed to be of the form
(\ref{j51}) with (\ref{j5}). The coefficients $\alpha_{({\rm
a})}$, \dots\ etc., are introduced there phenomenologically. When
this quantity is fixed, one can reconstruct the force
$\frak{R}^i_{({\rm a})}$ (acting on the plasma particle) using the
integral equation (\ref{check37}). We considered the forces of
three types. First, when ${\cal G}_{({\rm a})}{=}0$, one obtains
that $\frak{R}^i_{({\rm a})}{=}0$, and we deal with plasma
particles influenced by pure Lorentz force; this case relates to
the vanishing entropy production. Second, when $\frak{R}^i_{({\rm
a})} p_i \neq 0$, we deal with the model, in which the particle
mass is not constant and depends on the square of the pseudoscalar
field (\ref{check005}). We indicated the third version of the
force, for which $\frak{R}^i_{({\rm a})} p_i = 0$, as the
ponderomotive force; we illustrated the procedure of
reconstruction of this force with the example of
$\frak{R}^i_{({\rm a})}$ quadratic in the particle four-momentum
(see (\ref{check4})). We expect that further development of this
theory and (probably) new experimental data will clarify the
structure of this force.

%\vspace{3mm}
\appendix

\section{}\label{AppA}

Let us transform the pseudotensor
\begin{equation}
\Fst^{mn} \ F_{mq} = \frac{1}{2} \epsilon^{mnkl} F_{kl} F_{mq}
\label{1App1}
\end{equation}
using the standard decomposition of the Maxwell tensor
\begin{equation}
F_{kl} = E_k U_l - E_l U_k - \epsilon_{kljs} B^j U^s \,,
\label{1App2}
\end{equation}
where $E_l$ is the electric field four-vector,  $B^j$ is the
magnetic excitation four-vector, $U_k$ is the time-like velocity
four-vector of the observer. The four-vectors $E_l$ and $B^j$ are
orthogonal to the velocity four-vector, i.e.,  $E_l U^l =0 = B^j
U_j$. Taking into account the identity
\begin{equation}
\epsilon_{kljs} \epsilon^{kmnp} = - \delta^{mnp}_{ljs} \,,
\label{1App3}
\end{equation}
where $\delta^{mnp}_{ljs}$ is the six-indices Kronecker tensor defined as
\begin{equation}
\delta^{mnp}_{ljs} \equiv \delta^{m}_{l}\delta^{np}_{js} {+}
\delta^{m}_{s}\delta^{np}_{lj} {+} \delta^{m}_{j}\delta^{np}_{sl}
\,,\quad \delta^{np}_{js} =
\delta^{n}_{j}\delta^{p}_{s}{-}\delta^{n}_{s}\delta^{p}_{j} \,,
\label{1App4}
\end{equation}
we obtain by direct calculations that
\begin{equation}
\Fst^{mn} \ F_{mq} = \delta^n_q \ E_m B^m \,. \label{1App5}
\end{equation}
The convolution with respect to $n$ and $q$ yields
\begin{equation}
\Fst^{mn} \ F_{mn} =  4 E_m B^m \,, \label{1App6}
\end{equation}
thus the formula
\begin{equation}
\frac{1}{4} \delta^n_q \ \Fst^{mn} F_{mn} - \Fst^{mn} \ F_{mq} =0
\label{1App7}
\end{equation}
is valid. This relation can be also interpreted as follows: the
pseudotensorial analog of the stress-energy tensor of the
electromagnetic field is equal to zero identically.

\section{}\label{AppB}

In order to check the compatibility conditions in case of the
nonminimally extended model (see (\ref{EinBI})), we calculate
sequentially the divergences of all elements of the total
stress-energy tensor of the plasma, electromagnetic and
pseudoscalar fields nonminimally coupled to gravity. First of all,
we represent the divergence of the $T_{ik}^{(EM)}$ tensor as
follows:
\begin{gather}
\nabla^i T_{ik}^{(EM)}= {F_k}^n \nabla^m F_{nm} = F_{nk}
\left\{I^n  {+} \Fst^{nm} \nabla_m \phi + \nabla_m \left[{\cal
H}^{nm}+{\cal R}^{nmpq} F_{pq} {+} \phi {\chi}^{nmpq}_{({\rm A})}
\Fst_{pq}\right]\right\} \,,
\end{gather}
using the extended Maxwell equations (\ref{max2}) with (\ref{inducnm}). Then we transform the divergence of the tensor $T_{ik}^{(A)}$
\begin{gather}
\nabla^iT_{ik}^{(A)}=\Psi_0^2\left[\nabla^i\nabla_i\phi+m^2\phi+V'(\phi^2)\phi\right]\nabla_k
\phi ={}\nonumber\\
{}= {-} \nabla_k \phi \Biggl\{ \Psi^2_0 \left[ \nabla_m
\left(\Re^{mn}_{({\rm A})} \nabla_n \phi \right){+}\eta_{({\rm
A})} R\phi \right] +\sum_{({\rm a})} \int dP f_{({\rm a})}{\cal
G}_{({\rm a})} {+}
\frac{1}{4}\Fst_{mn}\left(F^{mn}{+}\chi^{ikmn}_{({\rm A})}F_{ik}
\right)\Biggr\} \,,
\end{gather}
using the nonminimal master equation for the pseudoscalar field
(\ref{eqaxi1}). From the sum of these two divergencies we extract
the terms linear in the parameter $q_1$ and compare it with the
divergence of the tensor $T_{ik}^{(1)}$:
\begin{equation}
    \nabla^i T_{ik}^{(1)}= - {F_k}^n \nabla^i(RF_{in}) \,.
\end{equation}
Then we continue this procedure, using the following formulas for the terms, which include $q_2$ and $q_3$:
\begin{equation}
    \nabla^i T_{ik}^{(2)}= - {F_k}^n \nabla^i(R_{il}{F^l}_{n}+F_{il}{R^l_n}) \,,
\end{equation}
\begin{equation}
    \nabla^i T_{ik}^{(3)} = - {F_k}^n \nabla^i(R_{inpq}F^{pq}) \,.
\end{equation}
When we deal with the term
\begin{equation}
\nabla^i{\cal T}_{ik}^{(1)} = -\nabla^i(R\phi \Fst_{in}){F_k}^n+\frac14RF_{pq}\Fst^{pq}\nabla_k\phi \,,
\end{equation}
we collect the expressions in front of the parameter
$(Q_1-\frac{1}{2}Q_{3})$, since we assume here that $Q_2=-Q_3$ and
take into account that ${\cal T}_{ik}^{(2)}$ can be transformed by
using (\ref{1App7}) and be included into the term
$\frac{1}{2}{\cal T}_{ik}^{(1)}$. Similarly, we compare the terms,
which contain $Q_3$, using
\begin{gather}
 \nabla^i{\cal T}_{ik}^{(3)}=-\frac12 {F_k}^n \nabla^i[(R_{inpq}-{}^{*}{R}^{*}_{inpq})\phi
 \Fst^{pq}] +\frac14R_{mnpq}F^{mn}\Fst^{pq}\nabla_k \phi \,,
\end{gather}
then compare the terms linear in $\eta_1$, $\eta_2$, $\eta_3$,
keeping in mind that
\begin{gather}
\nabla^i{\cal T}_{ik}^{(4)}=-\frac12 {F_k}^n
\nabla^i[(R_i^l\nabla_n\phi-R_n^l\nabla_i\phi)\nabla_l\phi]+
\frac12 \nabla_k\phi
\nabla^i[(R_{in}F^{ln}+F_{in}R^{ln})\nabla_l\phi]\,,\\
    \nabla^i{\cal T}_{ik}^{(5)}= \nabla_k\phi \nabla^i(R\nabla_i\phi)
    \,,\\
    \nabla^i{\cal T}_{ik}^{(6)}= \nabla_k\phi \nabla^i(R^n_i\nabla_n\phi) \,,
\end{gather}
and finally, the terms linear in $\eta_{{\rm A}}$, using the formula
\begin{equation}
    \nabla^i{\cal T}_{ik}^{(7)}=R\phi \nabla_k \phi \,.
\end{equation}
Direct calculations show that all the terms linear in the
mentioned coupling constants disappear, and the compatibility
conditions (\ref{EinBI}) written in the nonminimal case, reduce to
the form (\ref{4axi}) obtained for the minimal case. In other
words, the requirement (\ref{check37}) for the force-like
four-vector $\frak{R}^i_{({\rm a})}$ in the plasma nonminimally
coupled to gravity has the same form as in the minimal case, but
now the pseudoscalar ${\cal Q}_{({\rm a})}$ can contain the
appropriate nonminimal terms in addition to the terms written in
(\ref{j5}).

\begin{acknowledgments}
The work was partially supported by the Russian Foundation for
Basic Research (Grants Nos. 11-02-01162 and 11-05-97518), by the
Federal Targeted Program N14.T37.21.0668 and the State Assignment
N5.2971.2011.
\end{acknowledgments}


\begin{thebibliography}{99}


\bibitem{Vlasov1} Vlasov A A 1938 {\it JETP} {\bf 8} 291

\bibitem{Vlasov2} Vlasov A A 1968 {\it Sov. Phys. Usp.} {\bf 10} 721

\bibitem{MV1} Pitaevskii L P and Lifshitz E M 1981 {\it Physical Kinetics} (Butterworth-Heinenann Ltd, Oxford)

\bibitem{MV2} Piel A 2010 {\it Plasma Physics. An Introduction to Laboratory, Space and Fusion Plasmas} (Springer Verlag, Berlin, Heidelberg)

\bibitem{EV1} Rein G, Rendall A D and Schaeffer J 1995 {\it Commun. Math. Phys.} {\bf 168} 467

\bibitem{EV2} Andreasson H 2011 {\it Liv. Rev. Relat.} {\bf 14} 4

\bibitem{EVS1} Tegankong D 2005 {\it Class. Quantum Grav.} {\bf 22} 2381

\bibitem{EMV1} Ignat'ev Yu G and Balakin A B 1981 {\it Sov. Phys. J.} {\bf 24} 593

\bibitem{EMV4} Brodin G and Marklund M 1999 {\it Phys. Rev. Lett.} {\bf 82} 3012

\bibitem{EMV5} Andreasson H, Eklund M and Rein G 2009 {\it Class. Quantum Grav.} {\bf 26} 145003

\bibitem{FaraR} Faraoni V, Gunzig E and Nardone P 1999 {\it Fundam. Cosm. Phys.} {\bf 20} 121

\bibitem{HO} Hehl F W and Obukhov Yu N 2001 {\it Lect. Notes Phys.} {\bf 562} 479

\bibitem{OdinPhRep} Nojiri S and Odintsov S D 2011 {\it Phys. Rept.} {\bf 505} 59

\bibitem{derc1} Amendola L 1993 {\it Phys. Lett. B} {\bf 301} 175

\bibitem{derc2} Capozziello S and Lambiase G 1999 {\it Gen. Rel. Grav.} {\bf 31} 1005

\bibitem{Prasanna1} Prasanna A R 1971 {\it Phys. Lett. A} {\bf 37} 331

\bibitem{Drum} Drummond I T and Hathrell S J 1980 {\it Phys. Rev. D} {\bf 22} 343

\bibitem{Go} Goenner H F M 1984 {\it Found. Phys.} {\bf 14} 865

\bibitem{Kost1} Kostelecky V A and Mewes M 2002 {\it Phys. Rev. D} {\bf 66} 056005

\bibitem{Hehl2} Itin Ya and Hehl F W 2003 {\it Phys. Rev. D} {\bf 68} 127701

\bibitem{BL05} Balakin A B and Lemos J P S 2005 {\it Class. Quant. Grav.} {\bf 22} 1867

\bibitem{H3} Balakin A B and Zimdahl W 2005 {\it Phys. Rev. D} {\bf 71} 124014

\bibitem{BZ1} Balakin A B and Zayats A E 2007 {\it Phys. Lett. B} {\bf 644} 294

\bibitem{BSZ} Balakin A B, Sushkov S V and Zayats A E 2007 {\it Phys. Rev. D} {\bf 75} 084042

\bibitem{BDZ1} Balakin A B, Dehnen H and Zayats A E 2007 {\it Phys. Rev. D} {\bf 76} 124011

\bibitem{BaNi} Balakin A B and Ni W-T 2010 {\it Class. Quantum Grav.} {\bf 27} 055003

\bibitem{Rmatter1} Bertolami O, Lobo F S N and P\'aramos J 2008 {\it Phys. Rev. D} {\bf 78} 064036

\bibitem{Rmatter2} Puetzfeld D and Obukhov Yu N 2008 {\it Phys. Rev. D} {\bf 78} 121501

\bibitem{tide2} Zimdahl W and Balakin A B 1998 {\it Phys. Rev. D} {\bf 58} 063503

\bibitem{tide5} Zimdahl W, Schwarz D J, Balakin A B and Pav\'on D 2001 {\it Phys. Rev. D} {\bf 64} 063501

\bibitem{tide10} Balakin A B, Sussman R A and Zimdahl W 2004 {\it Phys. Rev. D} {\bf 70} 064027

\bibitem{A1} Perlmutter S J {\it et al} 1998 {\it Nature (London)} {\bf 391} 51

\bibitem{A2} Riess A G {\it et al} 1998 {\it Astron. J.} {\bf 116} 1009

\bibitem{A3} Schmidt B P {\it et al} 1998 {\it Astoph. J.} {\bf 507} 46

\bibitem{DE1} Copeland E J, Sami M and Tsujikawa S 2006 {\it Int. J. Mod. Phys.
D} {\bf 15} 1753

\bibitem{DE2} Frieman J, Turner M and Huterer D 2008 {\it Ann. Rev. Astron. Astrophys.} {\bf 46} 385

\bibitem{DE3} Padmanabhan T 2008 {\it Gen. Relat. Grav.} {\bf 40} 529

\bibitem{DM1} Del Popolo A 2007 {\it Astronomy Reports} {\bf 51} 169

\bibitem{DM2} Lazarides G 2007 {\it Lect. Notes Phys.} {\bf 720} 3

\bibitem{DM3} Silk J 2007 {\it Lect. Notes Phys.} {\bf 720} 101

\bibitem{Sp1} Battaner E and Florido E 2000 {\it Fund. Cosmic Phys.} {\bf 21} 1

\bibitem{Sp2} Sofue Y and Rubin V 2001 {\it Ann. Rev. Astron. Astrophys.} {\bf 39} 137

\bibitem{Peccei} Peccei R D and Quinn H R 1977 {\it Phys. Rev. Lett.} {\bf 38} 1440

\bibitem{8Ni} Ni W-T 1977 {\it Phys. Rev. Lett.} {\bf 38} 301

\bibitem{Weinberg} Weinberg S 1978 {\it Phys. Rev. Lett.} {\bf 40} 223

\bibitem{Wilczek0} Wilczek F 1978 {\it Phys. Rev. Lett.} {\bf 40} 279

\bibitem{Peccei2} Peccei R D 2008 {\it Lect. Notes Phys.} {\bf 741} 3

\bibitem{Battesti} Battesti R {\it et al} 2008 {\it Lect. Notes Phys.} {\bf 741} 199

\bibitem{Raffelt} Raffelt G G 1990 {\it Phys. Rep.} {\bf 198} 1

\bibitem{Turner} Turner M S 1990 {\it Phys. Rep.} {\bf 197} 67

\bibitem{Khl}  Khlopov M Yu and  Rubin S G 2004 {\it
  Cosmological Pattern of Microphysics in the Inflationary Universe}
(Kluwer Academic Publishers, Dordrecht)


\bibitem{BR1} Caldwell R R, Kamionkowski M and Weinberg N N 2003 {\it Phys. Rev. Lett.} {\bf 91} 071301

\bibitem{avo1} Capozziello S, De Laurentis M, Nojiri S and Odintsov S D 2009 {\it Phys. Rev. D} {\bf 79} 124007

\bibitem{LR1} Frampton P H, Ludwick K J and Scherrer R J 2011 {\it Phys. Rev. D} {\bf 84} 063003

\bibitem{19} Ni W-T 2008 {\it Prog. Theor. Phys. Suppl.} {\bf 172} 49

\bibitem{Wbook} Weinberg S 1972 {\it Gravitation and Cosmology} (John Wiley and Sons, New York)

\bibitem{HehlObukhov} Hehl F W and Obukhov Yu N 2003 {\it Foundations of classical electrodynamics: Charge, flux, and
metric} (Birkh\"auser, Boston)




\end{thebibliography}
\end{document}